\documentclass[review]{elsarticle}
\usepackage{lineno,hyperref}
\usepackage{tabls}
\usepackage{natbib}
\usepackage{epsf}
\usepackage{appendix}
\usepackage{ragged2e}
\usepackage{amsmath}
\usepackage{textcomp}
\usepackage{breqn}
\usepackage[top=1in, bottom=1.in, left=1.in, right=1.in]{geometry}
\usepackage{enumitem}
\setlist[itemize]{leftmargin=*}
\usepackage{caption}
\newcommand{\A}{{\cal A}}
\newcommand{\B}{{\cal B}}	
\newcommand{\weff}{{\omega_{\mathrm{eff},g}}}
\usepackage{lineno,hyperref}
\usepackage[table, dvipsnames]{xcolor}
\modulolinenumbers[5]

\journal{Journal of \LaTeX\ Templates}

\begin{document}

\begin{frontmatter}

\title{MULTI-GROUP DISCONTINUOUS ASYMPTOTIC \texorpdfstring{$P_1$}{Lg} APPROXIMATION IN RADIATIVE MARSHAK WAVES EXPERIMENTS}

\author{A.P. Cohen}
\ead{avnerco@gmail.com}
\author{S.I. Heizler}
\address{Department of Physics, Nuclear Research Center-Negev, P.O. Box 9001, Beer Sheva 8419001, ISRAEL}

%
%

\begin{abstract}
We study the propagation of radiative heat (Marshak) waves, using modified $P_1$-approximation equations. In relatively optically-thin media the heat propagation is supersonic,~i.e. hydrodynamic motion is negligible, and thus can be described by the radiative transfer Boltzmann equation, coupled with the material energy equation. However, the exact thermal radiative transfer problem is still difficult to solve and requires massive simulation capabilities. Hence, there still exists a need for adequate approximations that are comparatively easy to carry out. Classic approximations, such as the classic diffusion and classic $P_1$, fail to describe the correct heat wave velocity, when the optical depth is not sufficiently high. Therefore, we use the recently developed discontinuous asymptotic $P_1$ approximation, which is a time-dependent analogy for the adjustment of the discontinuous asymptotic diffusion for two different zones. This approximation was tested via several benchmarks, showing better results than other common approximations, and has also demonstrated a good agreement with a main Marshak wave experiment and its Monte-Carlo gray simulation. Here we derive energy expansion of the discontinuous asymptotic $P_1$ approximation in slab geometry, and test it with numerous experimental results for propagating Marshak waves inside low density foams. The new approximation describes the heat wave propagation with good agreement. Furthermore, a comparison of the simulations to exact implicit Monte-Carlo slab-geometry multi-group simulations, in this wide range of experimental conditions, demonstrates the superiority of this approximation to others.

\end{abstract}

\begin{keyword}
\texttt{Radiative transfer, Marshak waves, $P_1$ approximation}
\end{keyword}

\end{frontmatter}


\section{INTRODUCTION} 

The problem posed by the transfer of supersonic radiation (Marshak) waves in matter is of considerable importance in high energy density physics~\cite{castor2004,lindl}. Its solution greatly influences the ability to achieve inertial confinement fusion (ICF), and to describe radiation procedures in stars. This problem is modeled via the Boltzmann (transport) equation for photons, coupled to the matter via the energy balance equation~\cite{castor2004,Pomraning1973}. 

In the recent decades, several experiments involving the propagation of supersonic Marshak waves through low-density foams have been carried out and reported~\cite{Massen,BackPRL,Back2000,TWOP,TWOP2,Moore2015} (for a short review, see~\cite{CohenPRR}). Typically, in these experiments high energy laser beams are shot into hohlraums, which radiate an X-rays into a dilute foam attached to the hohlraum. These experiments help validate radiative heat theoretical models. As noted, the exact solution of the Boltzmann equation is the principal component of these models, beside microscopic opacity data.

The Boltzmann equation can be solved numerically by Implicit Monte-Carlo (IMC) simulations, which are exact when the number of histories goes to infinity~\cite{IMC}. Alternatively, a discrete ordinates ($S_N$) method can be used, which is exact when the number of ordinates goes to infinity, or the spherical harmonics ($P_N$) approximation, which is exact when the number of moments goes to infinity~\cite{Pomraning1973}. However, these calculations are still hard to carry out, especially in more than one dimension, and when scanning many physical parameters. Hence, there still exists a need for good adequate approximations that are comparatively easy to carry out~\cite{Olson1999}.

When the matter is optically thick the Boltzmann equation can be approximated by the diffusion equation~\cite{Pomraning1973}. However, when the number of mean free paths in the matter is close to one, the diffusion equation is no longer valid. The full $P_1$ equations, that give rise to the Telegrapher's equation, has a hyperbolic form, but with an incorrect finite velocity, $c/\sqrt{3}$~\cite{Heizler2010,HeizlerRavetto}. Possible solutions, such as flux-limiters (FL) solutions (in the form of a non-linear diffusion notation), or Variable Eddington Factor (VEF) approximations (in the form of full $P_1$ equations), yield a gradient-dependent nonlinear diffusion coefficients (or a gradient-dependent Eddington factor)~\cite{Olson1999,Su2001}, and are still not exact enough to be trusted to make a full reconstruction of experimental results~\cite{CohenJCTT}.

The {\em discontinuous asymptotic $P_1$ approximation} is being derived to approximate the Boltzmann solution in heterogeneous media. It rests on two foundations: (1) the asymptotic $P_1$ approximation~\cite{Heizler2010,Heizler2012,harel2020time,harel2020time2}; (2) the asymptotic jump conditions in the boundary between two different media, assuming flux continuity (and thus energy conservation), and a discontinuity in the energy density~\cite{zimmerman1979}. This approximation is based on an {\bf asymptotic thick limit approach}, however, due to the jump conditions, converging to the exact solution much faster than classic diffusion.  
These modifications allow the use of a $P_1$ form that is accurate even in highly-anisotropic scenarios. This approximation was tested via well-known benchmarks, showing better results than other common approximations~\cite{Cohen2018}. In addition, it was also compared to the results in one of the main Marshak wave experiments (using $\mathrm{SiO_2}$ foam~\cite{Back2000}) and IMC gray (mono-energetic) simulations of the same problem, showing good agreements~\cite{CohenJCTT}.

In the Marshak wave experiments, the radiation flux that breaks out from the end of the sample is measured in a specific energy channel. Hence, the gray solution is not sufficient for analyzing the experimental results, and we have to solve the complete multi-energy problem. In the present study we developed a {\bf non-gray} modification of the {\em discontinuous asymptotic $P_1$ approximation}, using multi-group notations. This approximation is validated by comparing it with numerous experiments, in different physical regimes that have recently been published.

\section{THE MULTI-GROUP DISCONTINUOUS ASYMPTOTIC \texorpdfstring{$P_1$}{Lg} EQUATIONS}
The radiative-transfer Boltzmann equation in the frequency-dependent case is:
\begin{equation}\label{Boltz}
\begin{split}
\frac{1}{c} & \frac{\partial I(\hat{\Omega},\vec{r},t,\nu)}{\partial t}+\hat{\Omega}\cdot\vec{\nabla}I(\hat{\Omega},\vec{r},t,\nu)+\left(\sigma'_{a}(\nu,T_m(\vec{r},t))+\sigma_{s}(\nu,T_m(\vec{r},t))\right)I(\hat{\Omega},\vec{r},t,\nu)=\\
&\sigma'_{a}(\nu,T_m(\vec{r},t)){B}(\nu,T_m(\vec{r},t))+\frac{\sigma_{s}(\nu,T_m(\vec{r},t))}{4\pi}\int_{4\pi}I(\hat{\Omega},\vec{r},t,\nu)d\hat{\Omega}+S(\hat{\Omega},\vec{r},t,\nu)\\
\end{split}
\end{equation}
where $I(\hat{\Omega},\vec{r},t,\nu)$ is the specific intensity of radiation at position $\vec{r}$ propagating in the $\hat{\Omega}$ direction at time $t$ and frequency $\nu$. $B(\nu,T_m)$ is the black-body radiation term, where $T_m(\vec{r},t)$ is the material temperature, $c$ is the speed of light and $S(\hat{\Omega},\vec{r},t)$ is an external radiation source.
$\sigma'_{a}$ and $\sigma_{s}$ are the absorption (opacity) and scattering cross-sections, respectively. Eq.~\ref{Boltz} assumes an elastic isotropic scattering (i.e. Thomson scattering). 
 
Along with the equation for the radiation energy, the complementary equation for the material is:
\begin{equation}\label{Matter1}
\frac{C_v(T_m(\vec{r},t))}{c}\frac{\partial T_m(\vec{r},t)}{\partial t}=
\sigma'_{a}(\nu,T_m(\vec{r},t))\left(\frac{1}{c}\int_{4\pi}{I(\hat{\Omega},\vec{r},t,\nu)d\hat{\Omega}}-aT_m^4(\vec{r},t)\right)
\end{equation}
where $C_v(T_m(\vec{r},t))$ is the heat capacity of the material and $a$ is the radiation constant ($aT_m^4=\int_{0}^{\infty}d\nu B(\nu,T_m)$).

Usually, the energy (frequency) dependency is modeled via the multi-group approximation~\cite{Pomraning1973}. In this approximation the energy space is divided into $G$ discrete groups, defining a group specific intensity $I_g(\hat{\Omega},\vec{r},t)=\int_{\nu_{g-1}}^{\nu_g}d\nu I(\hat{\Omega},\vec{r},t,\nu)$. We also defined $b_g$ coefficient as
$b_g=\int_{\nu_{g-1}}^{\nu_g}d\nu B(\nu,T_m)\Big/aT_m^4)$
, and the group Rosseland mean opacity:
\begin{equation}\label{sig_g}
\frac{1}{\sigma'_{ag}}=
  \int_{\nu_{g-1}}^{\nu_g}d\nu \frac{1}{\sigma'_{a}(\nu,T_m)}\frac{\partial B(\nu,T_m)}{\partial T_m}
  \Bigg/
  \int_{\nu_{g-1}}^{\nu_g}d\nu\frac{\partial B(\nu,T_m)}{\partial T_m}
\end{equation}
The energy-dependent Boltzmann equation (Eq.~\ref{Boltz}) is now replaced 
with coupled $G$ mono-energetic equations:
\begin{equation}
\frac{1}{c}\frac{\partial I_g(\hat{\Omega},\vec{r},t)}{\partial t}+\hat{\Omega}\cdot \vec{\nabla}I_g(\hat{\Omega},\vec{r},t)=\sigma'_{ag}(T_m(\vec{r},t))\left[b_g\frac{ac}{4\pi}T_m^4(\vec{r},t)-I_g(\hat{\Omega},\vec{r},t)\right]+S_g(\hat{\Omega},\vec{r},t)
\label{Boltz_g}
\end{equation}
In Eq.~\ref{Boltz_g} we ignore the scattering term since in the 100-300eV range, scattering is negligible. In the present study, we have used 100 groups opacity cross-sections tables 
using the CRSTA approximation~\cite{Kurz2012}. The foam heat capacities $C_v$ were taken from QEOS tables~\cite{QEOS}.

The first two angular moments of the group-dependent specific intensity $I_g(\hat{\Omega},\vec{r},t)$ can be expressed as:
\begin{equation}
E_g(\vec{r},t)=\frac{1}{c}\int_{4\pi}{I_g(\hat{\Omega},\vec{r},t,)d\hat{\Omega}}
\label{Edf}
\end{equation}
\begin{equation}
\vec{F_g}(\vec{r},t)=\int_{4\pi}{ I_g(\hat{\Omega},\vec{r},t)\hat{\Omega}d\hat{\Omega}}
\label{Fdf}
\end{equation}
where $E_g(\vec{r},t)$ is the group energy density, and $\vec{F_g}(\vec{r},t)$ is the group radiation flux. Integrating Eq.~\ref{Boltz_g} over all solid angle $\int{{d}\hat{\Omega}}$ yields the exact {\em conservation law}:
\begin{equation}
 \frac{1}{c}\frac{\partial E_g(\vec{r},t)}{\partial t}+
 \frac{1}{c} \nabla\cdot\vec{F_g}(\vec{r},t)=\sigma'_{ag}(T_m(\vec{r},t))\left(b_{g}acT_m^4(\vec{r},t)-E_g(\vec{r},t)\right)+\frac{S_g(\vec{r},t)}{c}
\label{Rad1}
\end{equation}
Integrating Eq.~\ref{Boltz_g} with $\int{\hat{\Omega{d}}\hat{\Omega}}$, assuming the classic $P_1$ closure (when the specific intensity is decomposed of its two first moments) yields:
\begin{equation}
 \frac{\A_g(\vec{r},t)}{c}\frac{\partial \vec{F_g}(\vec{r},t)}{\partial t}+
c\vec{\nabla}E_g(\vec{r},t)+\B_g(\vec{r},t)\sigma'_{ag}(T_m(\vec{r},t)) \vec{F}_g(\vec{r},t)=0
\label{AsympRad}
\end{equation}
when $\A_g=\B_g=3$. In the {\em{asymptotic $P_1$ approximation}} instead, the coefficients
$\A_g(\vec{r},t)$ and $\B_g(\vec{r},t)$ are media-dependent, which have closed sets of functions (not free parameters), of $\weff(\vec{r},t)$, the {\em group-dependent} mean number of particles that are emitted per collision or source terms and (for the non-scattering case) is defined by~\cite{Pomraning1973,Cohen2018}:
\begin{equation}
    \weff(\vec{r},t) = \frac{\sigma'_{ag} b_{g}acT_m^4(\vec{r},t) + S_g(\vec{r},t)/c}{\sigma'_{ag} E_g(\vec{r},t)}.
\label{omega}
\end{equation}
Notice that in the case of LTE, $\weff\equiv 1$, the asymptotic distribution tends to the diffusion limit. The coefficients are derived from an asymptotic derivation, both in space and time, as detailed in~\cite{Heizler2010,Heizler2012}. For the exact functional dependence of $\A_g(\weff)$ and $\B_g(\weff)$, please see~\cite{Heizler2010,Cohen2018}. We note that setting $\A_g=1$ and $\B_g=3$ reproduces the $P_{1/3}$ approximation, which yields the correct particle velocity~\cite{Olson1999}.

The asymptotic $P_1$ approximation supplies a good description in isotropic medium, especially in late times. It yields the exact steady-state asymptotic solution, similar to the \textit{the asymptotic diffusion approximation}~\cite{Case1953} and the (almost) correct particle velocity. However, in heterogeneous media, for example, in a system with sharp boundaries between media, this approximation is {\em not} good enough. 

A series of studies yield the exact solution of two adjacent semi-infinite half-spaces problem (the two-region Milne problem) yielding the exact boundary conditions (when both the asymptotic radiation flux and the asymptotic energy density are discontinuous)~\cite{mccormick1,mccormick2,mccormick3,ganapol_pomraning}. 
Zimmerman has offered a discontinuous asymptotic diffusion version,
based on Marshak-like approximation for the asymptotic jump conditions~\cite{zimmerman1979}. This approximation assumes continuous radiation flux $\vec{F}_A=\vec{F}_B$ (between semi-infinite medium $A$ and semi-infinite medium $B$) and discontinues energy density, $\mu_AE_A=\mu_BE_B$, when $\mu(\weff)$ is a medium dependent function, and derived by applying the Marshak boundary condition on the asymptotic distribution near a boundary between two different media~\cite{Pomraning1973}. We~\cite{Cohen2018} expanded these boundary jump conditions to the whole medium for the time-dependent asymptotic $P_1$ equation (Eq.~\ref{AsympRad}):
\begin{equation}  
\mu_g(\vec{r},t)\frac{\A_g(\vec{r},t)}{c}\frac{\partial \vec{F}_g(\vec{r},t)}{\partial t}+c\vec{\nabla}\left({\mu_g(\vec{r},t)}E_g(\vec{r},t)\right)+
\mu_g(\vec{r},t)\B_g(\vec{r},t){\sigma'_{ag}(T_m(\vec{r},t))}\vec{F}_g(\vec{r},t)=0
\label{DisC2}
\end {equation}
when we have added the index $g$ for the multi-group case.

The approximation has been tested versus the known \textbf{gray} Su-Olson Benchmark~\cite{SuOlson1996} and the nonlinear Olson benchmark~\cite{Olson1999}, showing better results than other common approximations~\cite{Cohen2018}. Although these benchmarks are characterized in highly-anisotropic flux profile, these modifications that are based on the exact asymptotic solutions, yield good results using $P_1$-form equations. In addition, 
a {\bf gray} comparison in one Marshak wave experiment (using $\mathrm{SiO_2}$ foam) showed a good agreement between the discontinuous asymptotic $P_1$ on the one hand, and the IMC and the experimental results on the other~\cite{CohenJCTT}.
The experimental result is of course non-gray and thus different, due to the specific measure in a specific energy band. In this work we expand the testing in fully multi-group treatment, in various of experimental setups.

\section{THE EXPERIMENTS} 
\label{sec:expdiscription}
\begin{figure}[htbp!]
\centering 
{
\includegraphics[width=10cm]{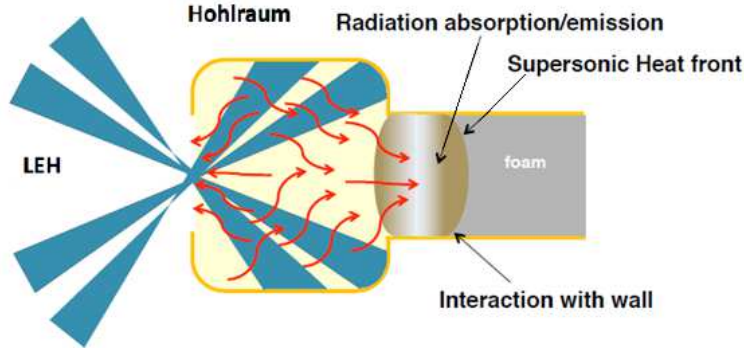}
}
\caption{A schematic diagram of typical Marshak wave experiments. Laser beam (blue lines) heats the hohlraum, which re-emits soft X-ray radiation (red arrows) into a dilute foam (gray), then, a Marshak wave propagates in the foam. The figure is taken from~\cite{MoorePresentation}.}  
\label{fig:schem1}
\end{figure}

The different experiments that are examined in this study all possess a common procedure, which is presented schematically in Fig.~\ref{fig:schem1}. The blue rays represent high energy (1kJ-10kJ) laser beams that are been shot into a small ($\sim$1-3mm) high-Z cavity (usually made of gold), i.e. hohlraum; the energy is absorbed in the hohlraum walls and heats them. Then, the walls re-emit a soft X-ray radiation (red arrows in Fig.~\ref{fig:schem1}) which, generates a heat wave that propagates inside low density foam. A high-Z (usually gold) cylinder coats the foam and prevents energy leakage from the foam's external surface.

When the heat wave front reaches the edge of the foam, the radiation flux is measured as a function of time by an X-ray steak camera or an X-ray diode (XRD)~\cite{BackPRL,Back2000,TWOP,Moore2015}. The radiation temperature in the hohlraum is measured as a function of time, usually through the laser entrance hall (LEH in Fig.~\ref{fig:schem1}). We use this temperature to estimate the heat wave drive temperature profile $T_D(t)$, assuming a black body radiation source (in frequency and direction) with a given drive temperature (for a wide discussion, see~\cite{CohenPRR,CohenJCTT}).
The main physical phenomena is the approximately one-dimensional (1D) slab-geometry heat wave propagation in the center of the cylinder. This phenomena is examined in the numerical simulations presented in the next section. Other effects, such as the energy leakage to the gold walls, the walls ablation and two-dimensional geometry effects, are of second importance, than the basic 1D phenomena~\cite{CohenPRR,CohenJCTT,Hurricane2006,McClarren2018} 

The properties of the different experiments that are examined in this study are summarized in Table~\ref{table:Optical length}. We specify the range of the optical depths (in mfp) in each experiment, which are reported in the each paper. In general, the range varies with the foam's length, i.e. shorter length means smaller optical depth. An exception is the Moore et al. experiment which has a given constant length, when the range of optical depth describes the thermodynamic track during the heat propagation through the foam~\cite{Moore2015}.
This parameter allow us to classify the limits of validity of the different approximations, comparing to the IMC solution and experimental results.
\begin{table}
\begin{center}
\begin{tabular}{||c | c | c | c | c ||} 
\hline
\rowcolor{GreenYellow}
{\bf The experiment} & {\bf Foam Type} & {\bf density} & {\bf Max. Temp.} & {\bf Optical} 
\\[0.5ex]
\rowcolor{GreenYellow}
 &  & {\bf [mg/cc]} & {\bf [eV]} &  {\bf depth}
\\[0.5ex] 
\hline
Back et al. PRL &  $\mathrm{SiO_{2}}$ &  10 & 85 & 1-2.5 \\
Back et al. POP &  $\mathrm{SiO_{2}}$ &  50 & 190 & 0.75-1.75 \\
Back et al. POP &  $\mathrm{Ta_2O_{5}}$ &  40 & 190 & 1.2-5 \\
Xu et al.  &  $\mathrm{C_6H_{12}}$, &  50 & 160 & 0.65-0.85   \\
Moore et al.  &  $\mathrm{C_8H_{7}Cl}$ &  100 & 310 & 2.5-7 \\
\hline
\end{tabular}
\end{center}
\caption{The different experiments studied in this paper. For each experiment, we specify the material of the foam, its density, and the maximal drive temperature and its approximated optical depth.}
\label{table:Optical length}
\end{table}

\section {BACK \texorpdfstring{$\mathrm{Ta_2O_5}$}{Lg} \& \texorpdfstring{$\mathrm{SiO_2}$}{Lg} HIGH- ENERGY EXPERIMENTS}
The first experiments we examine are the Back et al. $\mathrm{Ta_2O_5}$ and $\mathrm{SiO_2}$ high-energy experiments, which were carried out in the OMEGA-60 facility~\cite{Back2000}. The foam, either $\mathrm{SiO_2}$ at 50mg/cc or 40mg/cc $\mathrm{Ta_2O_5}$, was coated with a gold cylinder. Three different cylinder foam lengths were used for the $\mathrm{SiO_2}$ sample, and four different lengths in the $\mathrm{Ta_2O_5}$ case. The maximal drive energy temperature measured in these experiments was $\approx{190eV}$. The emitted radiation flux was measured by an X-ray streak camera, measuring photons in spectral band in about 550eV. The experimental results of the flux are shown in~\cite{Back2000} in arbitrary units.
\begin{figure}[htbp!]
\centering
(a)\includegraphics[width=7.5cm]{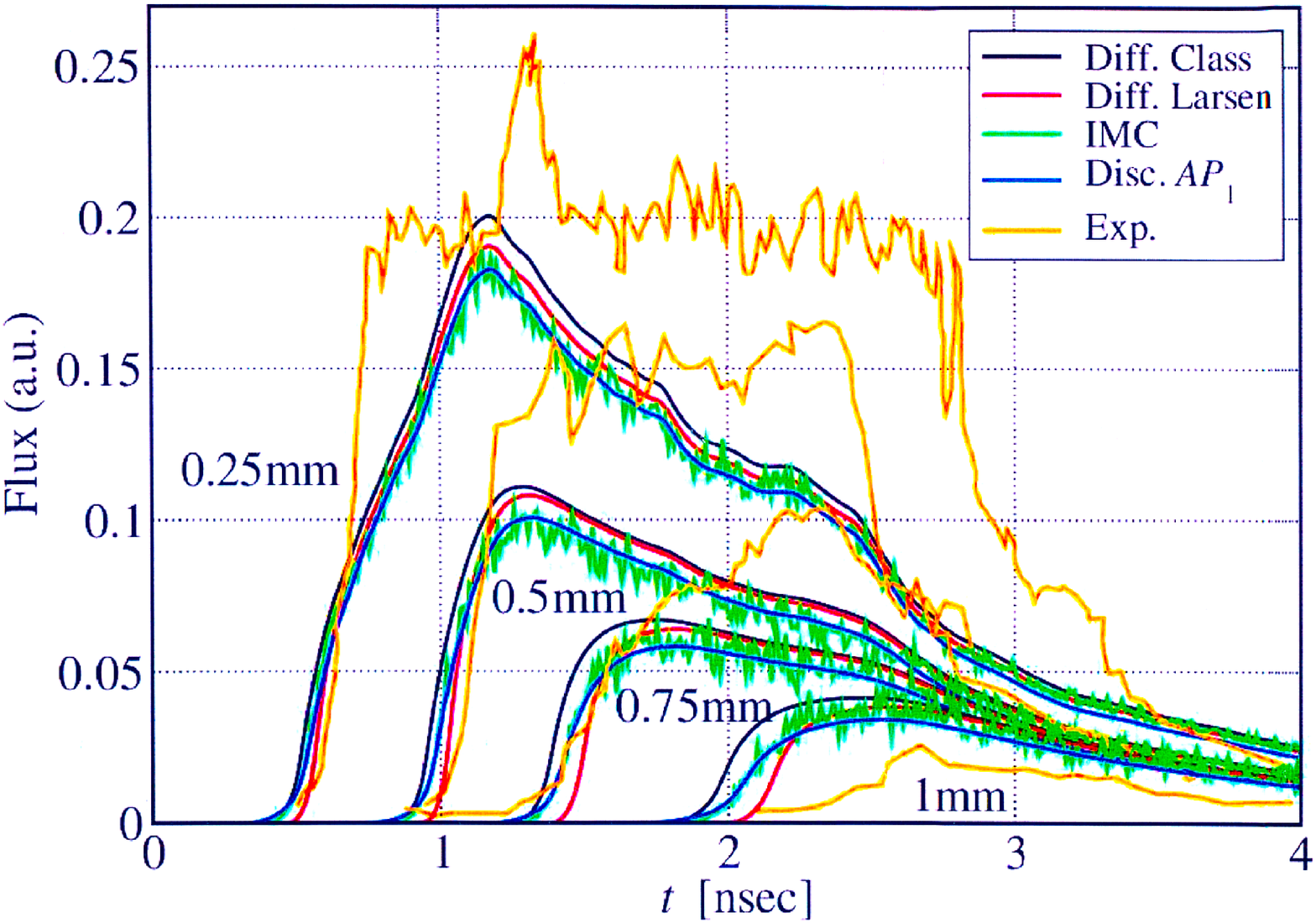}
(b)\includegraphics[width=7.5cm]{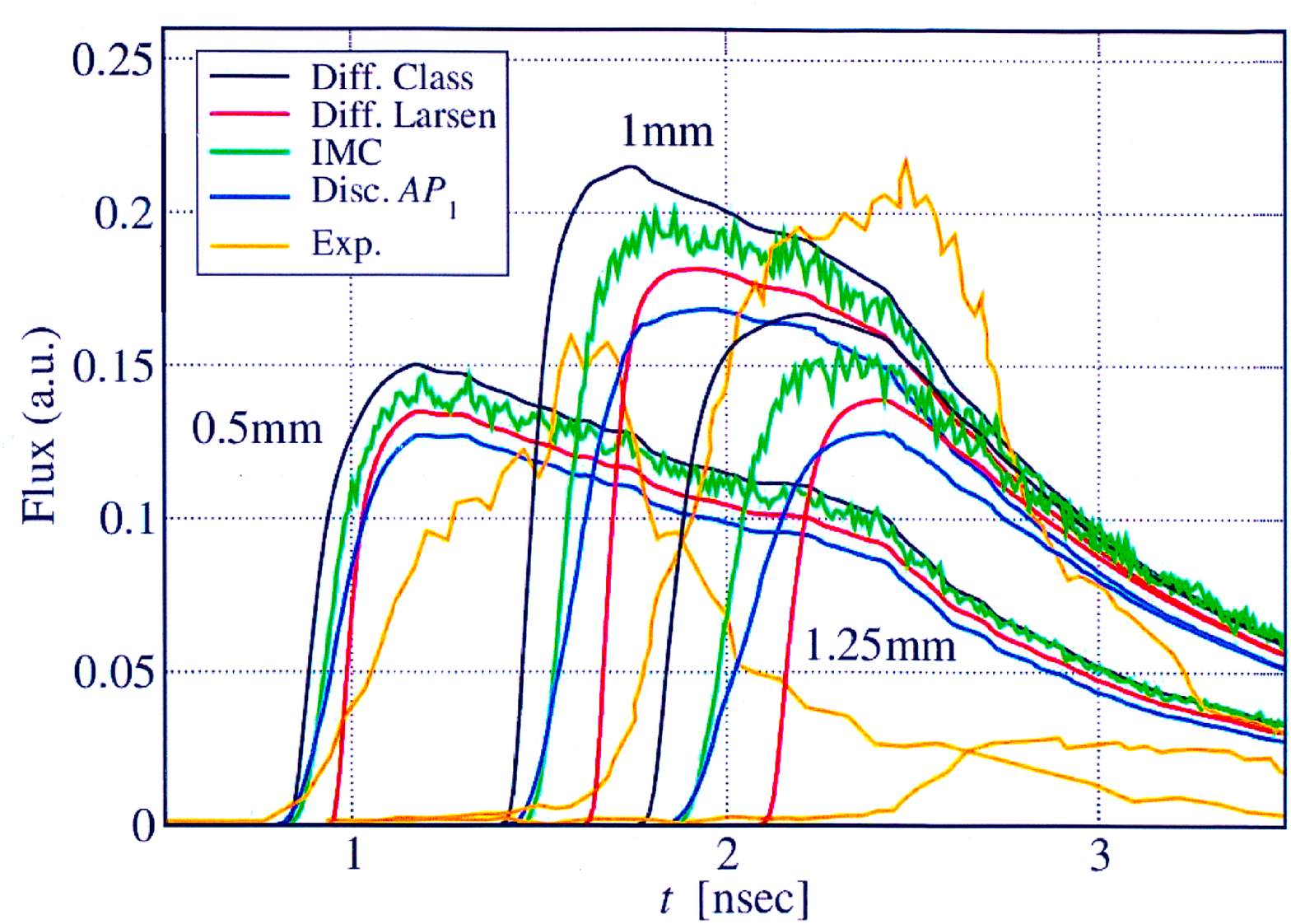}
\caption{
(a) The radiation flux that was emitted from the edge of the foam as function of time in Back et al. high energy $\mathrm{Ta_2O5}$ experiment~\cite{Back2000}. Four different experiments are represented at different foam lengths: 0.25, 0.5, 0.75, 1mm. (b) The $\mathrm{SiO_2}$ experiment~\cite{Back2000} is presented using three different foam lengths: 0.5, 1. 1.25 mm. The experimental results are represented by the orange curves. The green curves represent the 1D IMC simulations, the classic diffusion approximation in shown in black curves, the Larsen flux-limited diffusion in red and the discontinuous asymptotic $P_1$ approximation in blue curves.}
\label{fig:Back1}
\end{figure}

Fig.~\ref{fig:Back1} shows a comparison between the experimental results (orange), the 1D IMC solution (green) and different 1D approximations. The calculated flux in the simulation is in energy band of 535-580eV. The IMC numerical simulations units were scaled such that the IMC results will best fit to the experimental results in the 0.25mm long cylinder for the $\mathrm{Ta_2O_5}$ (Fig.~\ref{fig:Back1}(a)) and 1mm long in $\mathrm{SiO_2}$(Fig.~\ref{fig:Back1}(b)). All the other numerical simulations where scaled in the same factor. First, we note that in long foam lengths (1mm in $\mathrm{Ta_2O_5}$ and 1.25mm in $\mathrm{SiO_2}$), all simulations yield faster flux rise than do the experiments. This is due to 2D effects, that are broadly explained in~\cite{CohenPRR, CohenJCTT}.

It can be seen that the discontinuous asymptotic $P_1$ approximation (blue curves) fits the IMC solution nicely, especially in the $\mathrm{Ta_2O_5}$ experiment. In the $\mathrm{SiO_2}$ experiment, it yields very good agreement in the breakout times, but lower maximal results, due to the inherent discontinuity~\cite{Cohen2018}). Both the classic diffusion (black) and classic $P_1$ approximation (not shown) yields faster results, and the Larsen flux-limited diffusion~\cite{Olson1999} is late, especially in the $\mathrm{SiO_2}$ experiment. In general, the differences between the different approximations in the $\mathrm{Ta_2O_5}$ experiment are small, due to the relative high-opacity of this experiment (see Table.~\ref{table:Optical length}).

\section {BACK \texorpdfstring{$\mathrm{SiO_2}$}{Lg} LOW-ENERGY EXPERIMENT}

Another important experiment, also carried out in the OMEGA-60 facility, is the low-energy $\mathrm{SiO_2}$ experiment (the maximal drive energy temperature was $\approx 85eV$). Here, $\mathrm{SiO_2}$ foam at 10mg/cc density was coated with a gold cylinder. Three different experiments were carried out with three different foam lengths: 0.5, 1, 1.5mm. The emitted radiation flux was measured as a function of time with a spectral band in about $h{\nu}=$250eV energy.
\begin{figure}[htbp!]
\centering
  \includegraphics[width=7.5cm,angle=0.3]{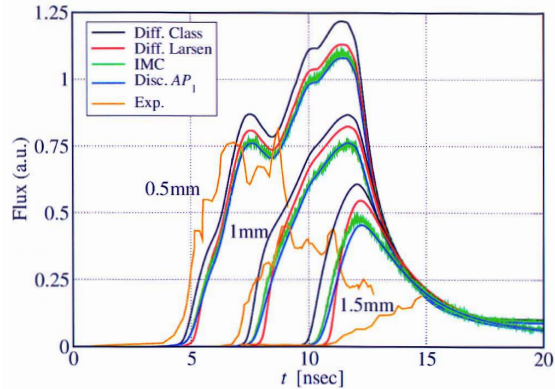}
\caption{ The radiation flux that was emitted out of the edge of the foam as function of time in the $\mathrm{SiO_{2}}$ low-energy experiment~\cite{BackPRL}. Three different foam lengths: 0.5, 1, 1.5 mm are presented.}  
\label{fig:BACK2}
\end{figure}

The experimental flux is presented in the orange curves in Fig.~\ref{fig:BACK2}. A 1D IMC simulation (green curves) demonstrates good agreement with the experimental result in the shorter lengths (0.5 and 1mm), especially in the break out times. The calculated flux in the simulation is in the spectral band of 200-297eV. The disagreement in the third length, 1.5mm, is due to the 2D effects, as mentioned before~\cite{CohenJCTT}. It can be seen that in this experiment, the discontinuous asymptotic $P_1$ approximation yields the closest agreement with the IMC simulations, by all means. This is due to the fact that this experiment is relatively optically thick, specifically in the larger lengths. In this experiment we have explicitly checked also the $P_{1/3}$ approximation, when the results were very close to the the classic diffusion approximation, as well classic $P_1$ approximation. This is not surprising since this problem is similar to the nonlinear Olson benchmark~\cite{Olson1999}.
\section {Xu EXPERIMENT}
\label{xu_sec}
\begin{figure}[htbp!]
\centering 
(a)\includegraphics[width=7.6cm]{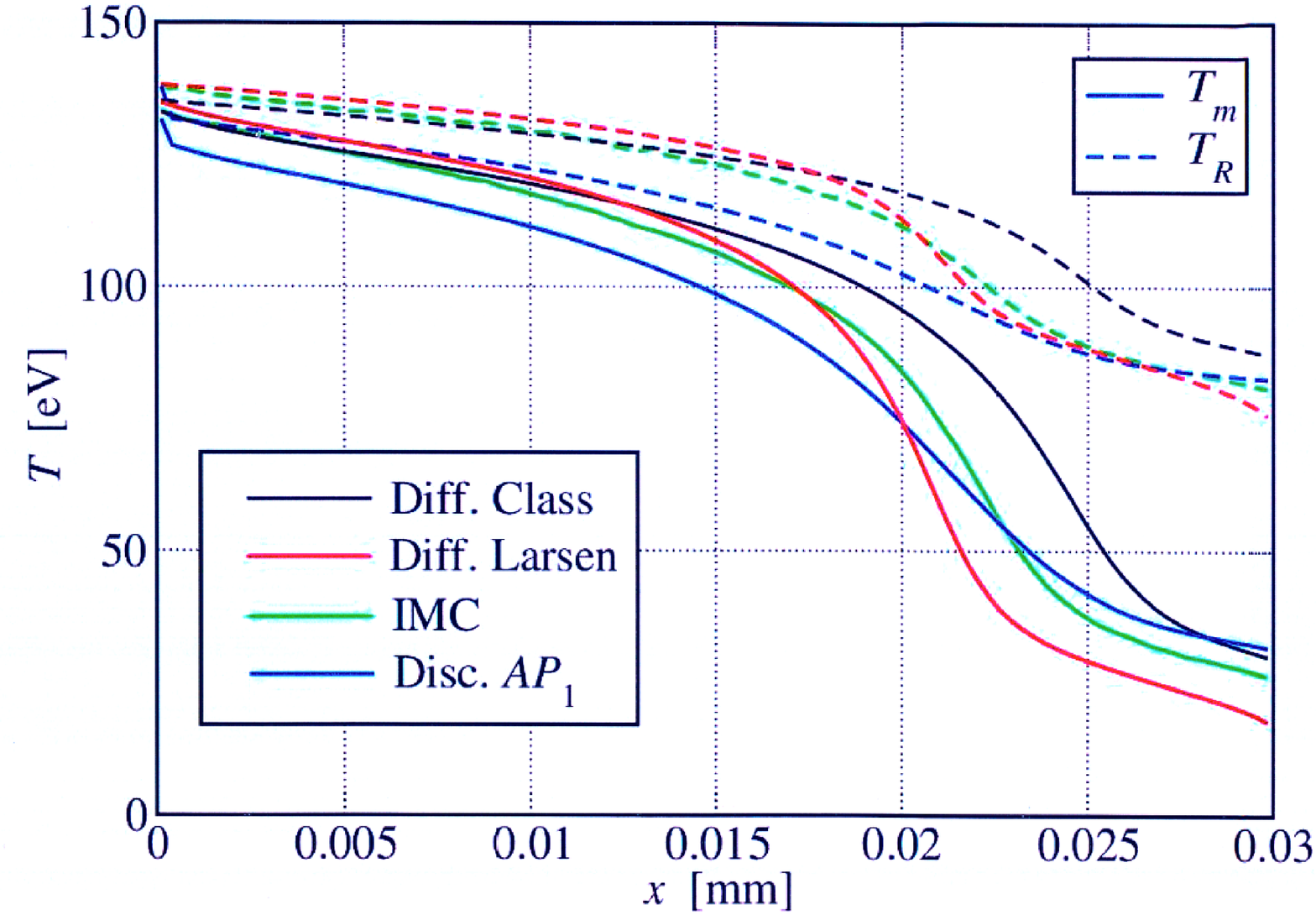}
(b)\includegraphics[width=7.4cm]{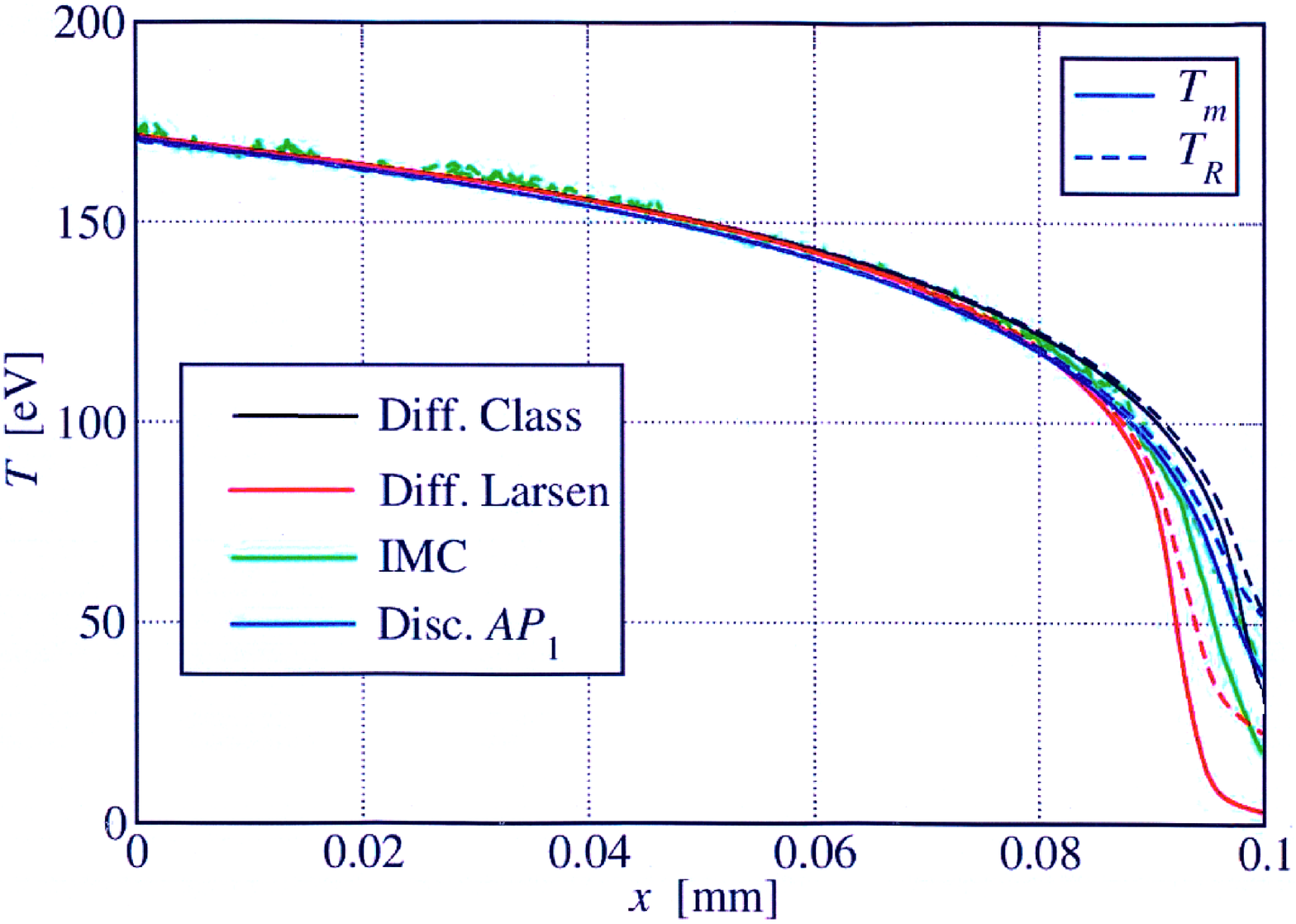}
\caption{(a) The $\mathrm{C_6H_{12}}$ experiment~\cite{TWOP} spatial temperature profiles are presented at time=0.9nsec. The green curves represent the 1D IMC simulations, the classic diffusion approximation in shown in black curves, the Larsen flux-limited diffusion in red and the discontinuous asymptotic $P_1$ approximation in blue curves. The material temperature, $T_m$ in the solid curves, and the effective radiation temperature, $T_r$ in the dashed curves. (b) The heat wave spatial temperature profiles, at time=1.9nsec in Back et al. high energy $\mathrm{Ta_2O5}$ experiment~\cite{Back2000}.}
\label{fig:HEAT_WAVE_PROFILE}
\end{figure}

The next experiment examined in this study was carried out in the SG-II facility~\cite{TWOP,TWOP2}. Cylinder foam made of 50mg/cc $\mathrm{C_6H_{12}}$ foam were used in two different lengths: 300 and 400$\mu$m. The maximal drive temperature inside the hohlraum was $\approx 160eV$. The CH foam is optically thin due to its low Z, in respect to the experiments been discussed earlier, with optical depth that is less than 1 (Table~\ref{table:Optical length}). Hence, the radiation flux starts to leak from the edge of the foam, before the material has been significantly heated (as can be seen later in Fig.~\ref{fig:HEAT_WAVE_PROFILE}(a)). In the multi group simulations that also cause the heat wave, spatial profile has a long "tail", and not a sharp front as in the gray simulations.
\begin{figure}[htbp!]
\centering 
(a)\includegraphics[width=7.7cm]{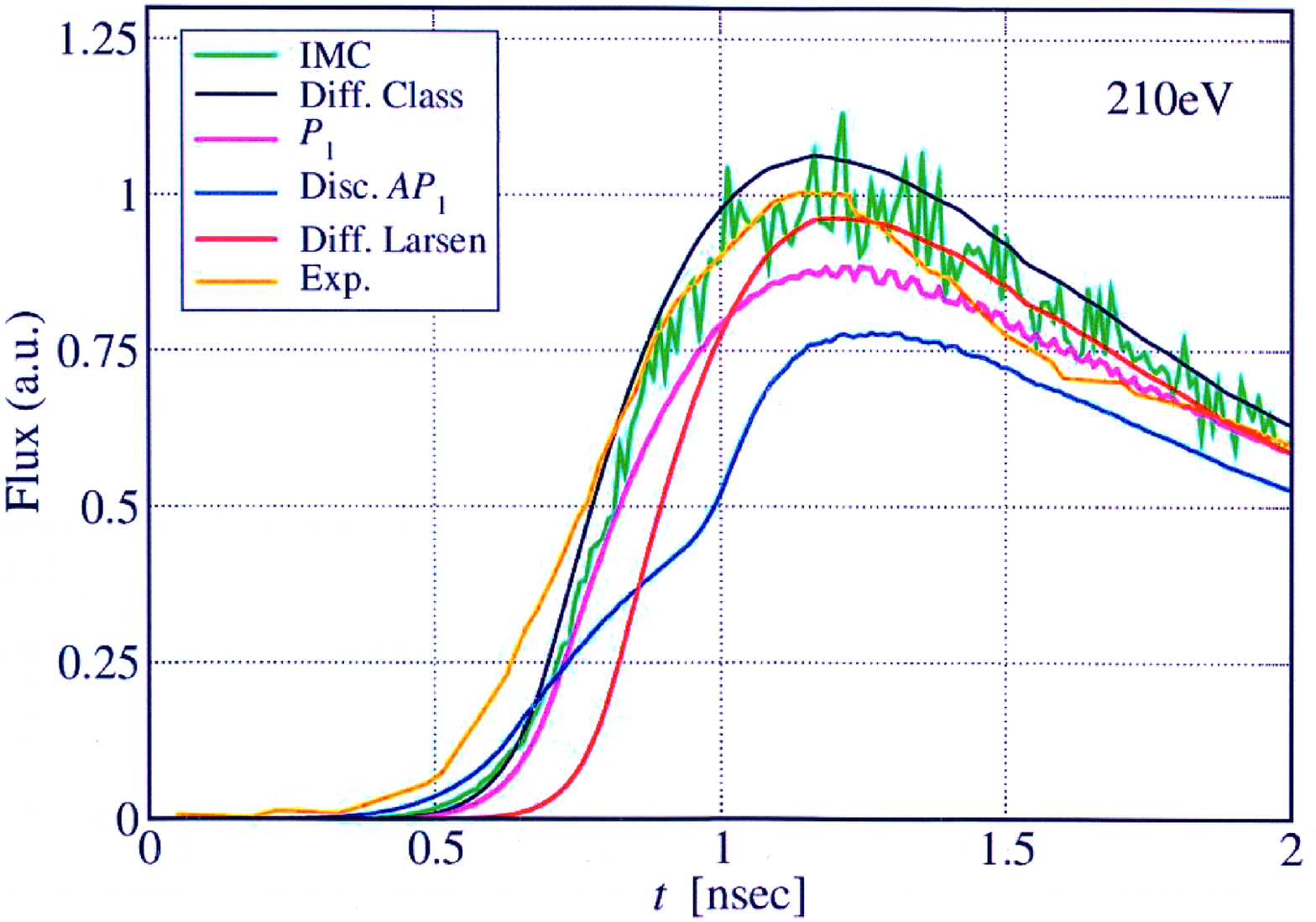}
(b)\includegraphics[width=7.2cm]{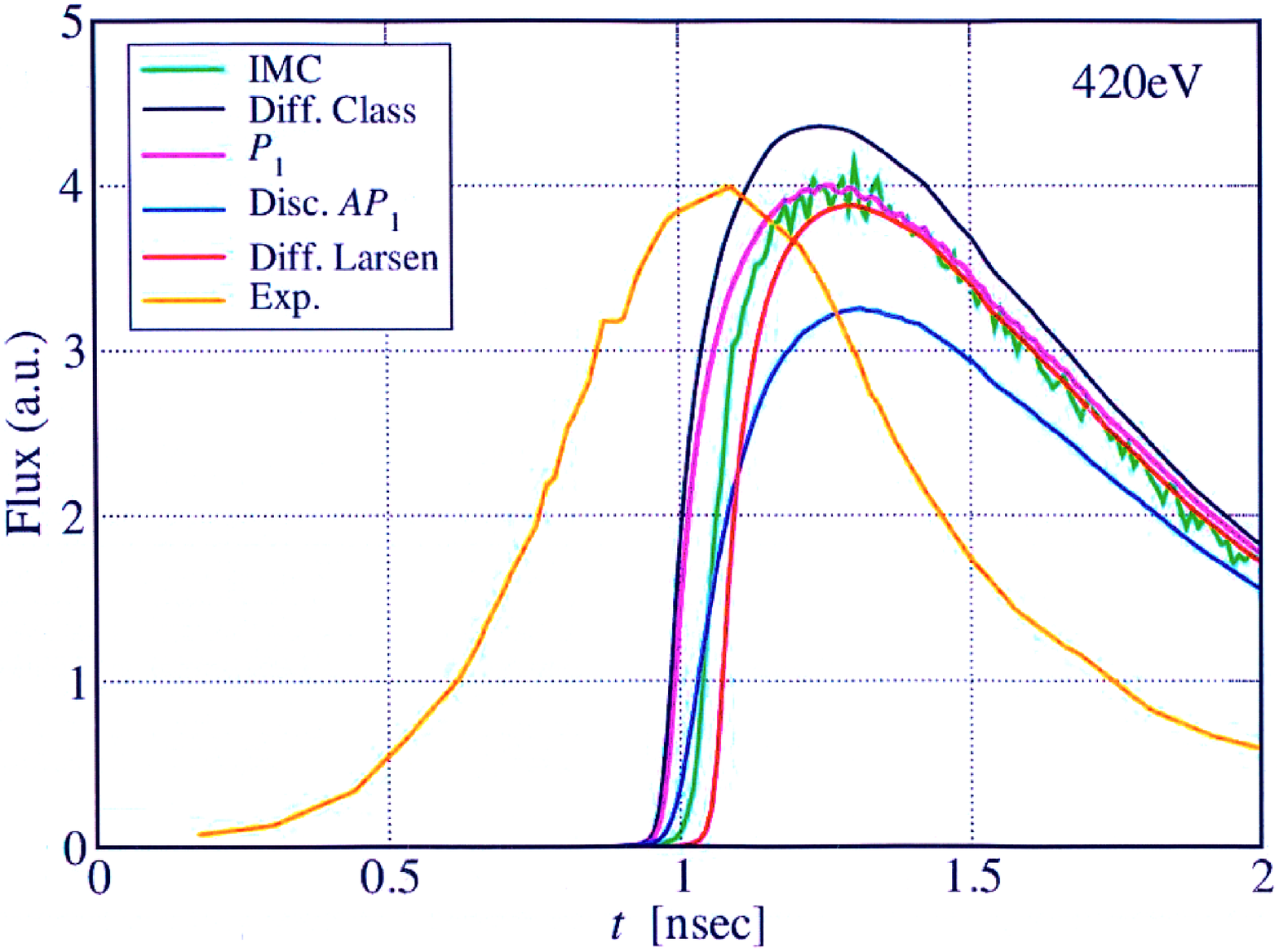}
\caption{A comparison between the experimental results (orange), and 1D simulations for the $\mathrm{C_6H_{12}}$ experiment~\cite{TWOP}, in two different energy bands with ($210eV$ - (a) and $420eV$ - (b)), in 300$\mu{m}$ foam length.}
\label{fig:Xu2}
\end{figure}

Two energy channels were used in the experiment to measure the flux coming out from the edge of the foam, in the 300$\mu{m}$ length foam. The first, around 210eV where the opacity is small, and around 420eV where the opacity is higher. The simulations presented in that paper~\cite{TWOP}, showed a significant difference between the two channels (though the experimental data is less decisive). This also demonstrates the importance of using multi-group simulations (gray simulations do not represent the real experimental picture). The earlier break out in the 210eV channel, is also shown in our IMC simulation shown in Fig.~\ref{fig:Xu2}(a) (green line), when the calculated flux is in energy band 169-223eV and in Fig.~\ref{fig:Xu2}(b) is in 365-468eV energies.

\begin{figure}[htbp!]
\centering 
\includegraphics[width=7.7cm]{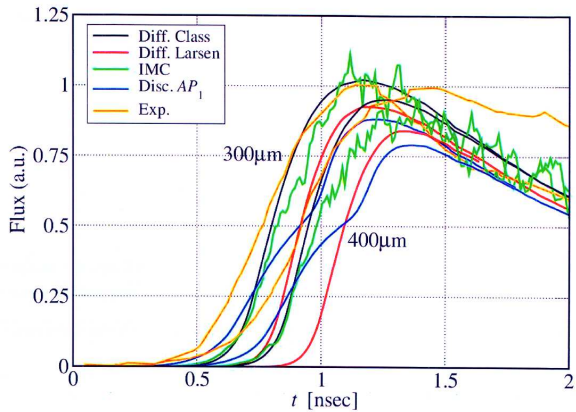}
\caption{A comparison between the experimental results (orange), and 1D simulations for the $\mathrm{C_6H_{12}}$ experiment~\cite{TWOP}, in two different foam length ($300\mu m$ and $400\mu m$).}
\label{fig:Xu1}
\end{figure}

A comparison of the radiated flux in the different foam lengths between the experimental results (orange), the 1D IMC solution (green) and the 1D approximations is shown in Fig.~\ref{fig:Xu1}. Here, as shown in both Fig.~\ref{fig:Xu2} and Fig.~\ref{fig:Xu1}, the classic diffusion (black) yields the closest agreement with the IMC simulations and the experimental results. The Larsen flux-limited diffusion (red) is significantly slower than the IMC simulations. Our discontinuous asymptotic $P_1$ approximation (blue) yields the correct breakout time (when the flux rises), but does not fit well at the maximum flux area. However, the relative gap between the two lengths is similar in all approximations and in the experiments. The low opacity of this experiment (less than 1 mfp) is the reason that the discontinuous asymptotic $P_1$ is less successive in this experiment, since it is an {\em asymptotic thick limit approach} approximation.

To demonstrate the low opacity of this experiment we show in Fig.~\ref{fig:HEAT_WAVE_PROFILE}(a) that the heat wave front (in $T_m$, the solid line) does not reach to the edge of the foam, at time=0.9nsec while the effective radiation temperature, $T_r=(E/a)^{1/4}$ has reached the edge at that time. In this optically thin medium, there are ``holes" of very optically thin energy bands. That mean that the flux that is measured in the 210eV range, and arises at 0.5-0.9nsec, does not represent the main heat wave progress. Note that the average radiation front profile in the discontinuous asymptotic $P_1$ is very similar to the IMC one (which explains the good agreement between IMC and the discontinuous asymptotic $P_1$ {\em in the breakout times}).

This difference, between $T_m$ and $T_r$ is because the heat wave transfers in an optically thin material. For comparison, we compare the heat wave spatial profile close to the break out time in the optically thick $\mathrm{Ta_2O_5}$ experiment (1.9nsec in 1mm foam length) in the different approximations (Fig.~\ref{fig:HEAT_WAVE_PROFILE}(b)). It can be seen that in this case, the effective radiation temperature, is very close to the material temperature, $T_m$, which means the system is close to local thermodynamic equilibrium. Again, the discontinuous asymptotic $P_1$ approximation (blue curves) heat wave front progresses at almost the same speed as in the IMC instance. .

\section {MOORE \texorpdfstring{$\mathrm{C_8H_7Cl}$}{Lg} EXPERIMENT}
The last experiment examined in this study is the most advanced and detailed set of experiments reported up to date, and are known as the Pleiades experiments~\cite{Moore2015}. These experiments were conducted in the high-power NIF facility and the drive temperature had reached $\approx{300}$eV shown in~\cite{Moore2015}. The experiment was carried out mostly with $\mathrm{C_8H_7Cl}$ foam (The Cl plays a major role in determining the foam opacity) 
in different densities (of about $\mathrm{100mg/cc}$).

In this study we examine one representative example, a $\mathrm{100mm/cm^3}$ $\mathrm{C_8H_7Cl}$ foam. The shot's energy was 367.3kJ, and the radiation incident flux to the foam was calculated numerically based on the measured flux radiated from the back of the hohlraum~\cite{Moore2015}. The 2.8mm long and 2mm diameter cylindrical foam was coated with Au tube. The 1D simulations cannot describe the experimental results because this experiment has significant 2D effects~\cite{CohenPRR}. However the 1D simulations still important for estimating the sensitivity to local variables, such as the exact opacity of the foam. Thus, we compare the different approximations to a 1D IMC simulations.
\begin{figure}[htbp!]
\centering
  \includegraphics[width=7.5cm]{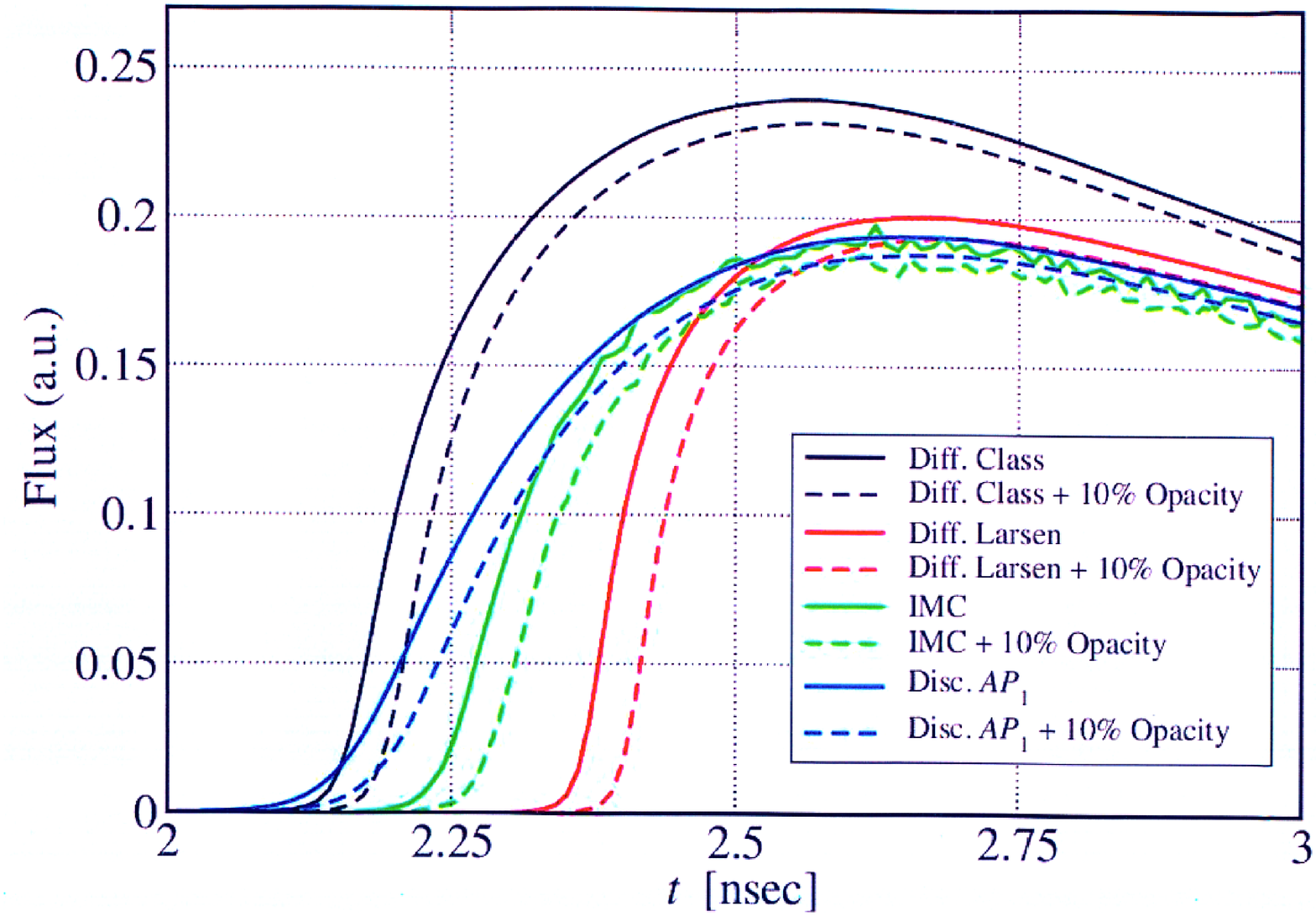}
\caption{A comparison between the 1D simulations for the $\mathrm{C_8H_{7}Cl}$ experiment, in two different opacity tables (The numerical CRSTA table, and the same opacity multiplied by 1.1 (dash curve).}
\label{fig:moore2}
\end{figure}

A comparison between the 1D IMC solution (green) and the 1D approximations is shown in Fig.~\ref{fig:moore2}, where the calculated flux in the simulation is in energy band of 80-969eV (similar to the energy band that was in use in the experiment). We can see that the classic diffusion (black) yields results that are faster and higher than the IMC results. The Larsen flux-limited diffusion (red) is significantly slower and lower than the IMC simulations. The discontinuous asymptotic $P_1$ Approximation (blue) yields the closest agreement with the IMC simulations. Again, this is because this experiment was conducted in a relatively optically-thick medium (see Table~\ref{table:Optical length}).

The nominal calculations were carried out with the nominal CRSTA table~\cite{Kurz2012} for $\mathrm{C_8H_7Cl}$. It is quite reasonable to assume an error bar of $\mathrm{10}\%$. Hence, we set the same calculations, only with a multiplied 1.1 CRSTA opacity factor (dash curves). The result points out that the discontinuous asymptotic $P_1$ Approximation, also estimates this effect correctly.

\section {QUANTITATIVE ANALYSIS}

To quantify the level of agreement of the different approximations, we summarize different parameters of the measured fluxes signals: The breakout (burnthrough) times (Table~\ref{table:Breaktimes}), and the maximal value of the signal (Table~\ref{table:MaxValue}), comparing to 1D IMC results.
The errors of the different approximations are written in parentheses, while the approximation with the smallest error is marked in red, in each experiment. The time at which the flux starts to rise is termed the breakout time, which we define as the time the flux reaches $\mathrm{25}\%$ of its maximal value. The maximal IMC flux values were calculated after smoothing (moving average with span 5-9, in different experiments). 

\begin{table}
\begin{center}
\begin{tabular}{||c | c | c | c | c | c ||} 
\hline
\rowcolor{SkyBlue}
Experiment Name & Foam's length &  Monte-Carlo & Diffusion & Flux-limited & Discontinuous \\
\rowcolor{SkyBlue}
Foam & [mm] &   &   &  diffusion & asymptotic $P_1$ \\
[0.5ex] 
\hline
Back PoP & 0.5 & 0.903 & 0.866 (-4.0\%) & 0.974 (7.9\%) & \textcolor{red}{0.903 (0.02\%)}\\
$\mathrm{SiO_2}$ & 1 & 1.533 & 1.444 (-5.8\%) & 1.664 (8.5\%)& \textcolor{red}{1.536 (0.17\%)}\\ 
 & 1.25 & 1.953 & 1.832 (-6.2\%) & 2.136 (9.4\%)& \textcolor{red}{1.963 (0.53\%)}\\ 
\hline
Back PoP & 0.25 & 0.573 & \textcolor{red}{0.569 (-0.75\%)}& 0.597 (4.2\%)& 0.584 (2.0\%) \\
$\mathrm{Ta_2O_5}$ & 0.5 & 0.983 & 0.957 (-2.6\%)& 1.024 (4.2\%)&  \textcolor{red}{0.984 (0.10\%)}\\ 
 & 0.75 & 1.403 & 1.370 (-2.3\%)& 1.485 (5.9\%)&  \textcolor{red}{1.404 (0.05\%)}\\ 
 & 1 & 2.023 & 1.950 (-3.6\%)& 2.129 (5.2\%)&  \textcolor{red}{2.002 (-1.0\%)}\\ 
\hline
Back PRL & 0.5 & 5.733 & 5.506 (-4.0\%)& \textcolor{red}{5.736 (0.06\%)}& 5.804 (1.2\%) \\
$\mathrm{SiO_2}$ & 1 & 8.103 & 7.756 (-4.3\%)& 8.336 (2.9\%)&\textcolor{red} {8.233 (1.6\%)}\\ 
 & 1.5 & 10.683 & 10.293 (-3.6\%)& 11.066 (3.6\%)&\textcolor{red} {10.800 (1.1\%)}\\ 
\hline
Xu & 0.3 & 0.734 & \textcolor{red}{0.723 (-1.5\%)}& 0.834 (13.6\%)& 0.693 (-5.6\%)\\
$\mathrm{C_6H_{12}}$ & 0.4 & 0.874 & \textcolor{red}{0.872 (-0.17\%)} & 1.004 (14.9\%)& 0.824 (5.7\%)\\ 
\hline
Moore & 2.8  & 2.263 & 2.172 (-4.0\%)& 2.373 (4.9\%)& \textcolor{red}{2.193 (-3.1\%)}\\
$\mathrm{C_8H_7Cl}$ & 2.8 + & 2.303 & 2.203 (-4.3\%)& 2.412 (4.8\%) & \textcolor{red}{2.233 (-3.0\%)}\\ 
  & 10\% Opacity &   &   &   &   \\ 
\hline
\end{tabular}
\par
\end{center}
\caption{Breakout times [nsec] of the different experiments studied in this paper. For each experiment, we compare the breakout times (flux reaches $\mathrm{25}\%$ of its maximum value) in the IMC simulation with the different approximations. The smallest difference is marked in red.}
\label{table:Breaktimes}
\end{table}

\begin{table}
\begin{center}
\begin{tabular}{||c | c | c | c | c | c ||} 
\hline
\rowcolor{GreenYellow}
Experiment Name & Foam's length &  Monte-Carlo & Diffusion & Flux-limited & Discontinuous \\
\rowcolor{GreenYellow}
Foam & [mm] &   &   &  diffusion & asymptotic $P_1$ \\
[0.5ex] 
\hline
Back PoP & 0.5 & 0.143 & 0.150 (5.4\%) & \textcolor{red}{0.135 (-5.4\%)}& 0.127 (-10.8\%)\\
$\mathrm{SiO_2}$ & 1 & 0.196 & 0.215 (10.0\%)&\textcolor{red}{ 0.182 (-7.1\%)}& 0.169 (-13.8\%)\\ 
 & 1.25 & 0.151 & 0.167 (10.4\%)& \textcolor{red}{0.139 (-8.1\%)}& 0.128 (-15.3\%)\\ 
\hline
Back PoP & 0.25 & 0.184 & 0.201 (9.1\%)& 0.190 (3.6\%)& \textcolor{red}{0.183 (-0.56\%)} \\
$\mathrm{Ta_2O_5}$ & 0.5 & 0.100 & 0.111 (10.5\%)& 0.108 (7.6\%)& \textcolor{red}{0.101 (0.38\%)}\\ 
 & 0.75 & 0.061 & 0.067 (10.1\%)& 0.064 (5.1\%)& \textcolor{red}{0.058 (-4.1\%)}\\ 
 & 1 & 0.039 & 0.041 (6.6\%)& \textcolor{red}{0.038 (-1.4\%)}& 0.034 (-12.0\%)\\ 
\hline
Back PRL & 0.5 & 1.105 & 1.219 (10.4\%)& 1.132 (2.5\%)& \textcolor{red}{1.081 (-2.1\%)} \\
$\mathrm{SiO_2}$ & 1 & 0.761 & 0.868 (14.0\%)& 0.824 (8.2\%)& \textcolor{red}{0.762 (0.07\%)}\\ 
 & 1.5 & 0.481 & 0.608 (26.4\%)& 0.546 (13.6\%)& \textcolor{red}{0.454 (-5.6\%)}\\ 
\hline
Xu & 0.3 & 1.046 & \textcolor{red}{1.022 (-2.3\%)}& 0.927 (-11.4\%)& 0.881 (-15.7\%)\\
$\mathrm{C_6H_{12}}$ & 0.4 & 0.899 & \textcolor{red}{0.951 (5.8\%)} & 0.841 (-6.4\%)& 0.790 (-12.1\%)\\ 
\hline
Moore & 2.8  & 0.193 & 0.240 (24.3\%)& 0.201 (4.0\%)& \textcolor{red}{0.194 (0.54\%)}\\
$\mathrm{C_8H_7Cl}$ & 2.8 + & 0.185 & 0.232 (25.0\%)& 0.193 (4.3\%)& \textcolor{red}{0.188 (1.2\%)}\\ 
  & 10\% Opacity &   &   &   &   \\ 
\hline
\end{tabular}
\par
\caption{Maximal flux values (a.u.) of the different experiments studied in this paper. For each experiment, we compare the maximal flux values in the IMC simulation with the different approximations. The smallest difference is marked in red.}
\label{table:MaxValue}
\end{center}
\end{table}

First, in Table~\ref{table:Breaktimes} the {\textit{discontinuous asymptotic $P_1$ approximation}} yields better results than the other approximations in all experiments - except the Xu experiment, which is extremely optically thin (Table~\ref{table:Optical length}), as explained (see Sec.~\ref{xu_sec}). Finally, checking the maximal flux values in Table~\ref{table:MaxValue}, the discontinuous asymptotic $P_1$ yields the best approximation in the opaque experiments: the Moore experiment, the $\mathrm{Ta_2O_5}$ experiment and the low-energy $\mathrm{SiO_2}$ Back experiment. In the optically thin experiments, i.e. the Xu and the high-energy $\mathrm{SiO_2}$ experiments, it yields maximal flux results that are too low. 

\section{CONCLUDING REMARKS}

Radiative heat (Marshak) waves experiments have been studied for the last three decades, and were carried out in the world's leading high energy laser facilities, such as OMEGA-60, SG-II and the NIF facilities. The basic physical process that is tested is the radiative transfer of thermal X-ray photons inside media of several Rosseland mean free paths, generating Marshak waves, as well as the microscopic opacities (that determines the mfp). 

The main equation that governs this physical process is the transport (Boltzmann) equation for photons. An exact solution for this equation is a heavy computation task and hard to obtain, thus good approximations may be very useful. Unfortunately, the classic well-known LTE diffusion approximation yields insufficient results, due to the relative low-number of mfp that characterizes these experiments (see Table~\ref{table:Optical length}). Recently, we 
have derived a new modified $P_1$ approximation, called the {\textit{discontinuous asymptotic $P_1$ approximation}}~\cite{Cohen2018,CohenJCTT}, which rests on some basic foundations of transport theory, such as the Case et al. asymptotic solution for infinite homogeneous transport equation~\cite{Case1953}, and the jump conditions of the asymptotic solution of the two-region Milne problem~\cite{mccormick1,mccormick2,mccormick3}, with no free parameters. This approximation, which was tested in
simple gray benchmarks, also yields many of the transport properties found in the exact full transport solution.

In this study we have expanded the discontinuous asymptotic $P_1$ approximation for multi-group conditions, which is essential for calculating the Marshak wave experiments, where the heat front is tracked via the measure of the radiated flux in specific energy-bands. The new approximation was validated by various experiments in different physical conditions. The agreement of the new approximation results with the IMC simulations in most of the experiments is surprisingly high. As this approximation rests on asymptotic behaviors of the exact transport equation, i.e. uses an {\bf asymptotic thick limit approach}, it yields better results as the optical depth of the experiment is larger. The Xu et al. experiment~\cite{TWOP2}, when the approximated mfp is less than 1, is an exception, when the new approximation yields results that are less satisfactory than other approximations. In those experiments not affected by 2D effects (the shorter lengths), the simulations also usually yield close agreement with the experimental results.

This study integrates the basic theoretical derivation of an approximation for the transport equation using asymptotic regimes, for modeling complicated Marshak waves experiments. The approximation that is based on basic milestone works in transport, and was first tested in analytic benchmarks, ultimately managed to describe full experimental results. In future studies we plan to examine whether the computational benefits to be derived from using simple modified Pi equations, as compared to full transport approaches, are more significant in multi-dimensions and in complicated radiation energy-flow scenarios. 

\section*{ACKNOWLEDGEMENTS}
We would like to thank Yonatan Elbaz for taking the time to read and comment on this paper. We acknowledge the support of the PAZY Foundation under Grant \textnumero~61139927.

\bibliography{bibliography}
\bibliographystyle{elsarticle-num}

\end{document}